\documentclass[pra,twocolumn,longbibliography,amsmath,superscriptaddress,amssymb]{revtex4-1}
\usepackage{graphicx}
\usepackage{dcolumn}
\usepackage{bm}
\usepackage{amssymb}
\usepackage{color}
\usepackage[dvipsnames]{xcolor}
\usepackage{placeins}
\usepackage{epstopdf}

\usepackage[normalem]{ulem}

\usepackage[linkcolor=blue,urlcolor=blue,citecolor=blue,colorlinks=true]{hyperref}

\newcommand{\spr}{\shortparallel}
\newcommand{\rtm}[1]{\mathrm{#1}}
\newcommand{\ai}{{\it ab initio}}
\newcommand{\kp}{\hbox{$\mathbf{k}\cdot\mathbf{p}$}}

\begin{document}

\title{Spatial aspects of spin polarization of structurally split surface states in thin films with magnetic exchange and spin-orbit interaction}

\author{I. A. Nechaev}
\affiliation{Donostia International Physics Center (DIPC), Paseo Manuel de Lardizabal 4, 20018 Donostia/San Sebasti\'{a}n,  Basque Country, Spain}
\affiliation{Department of Electricity and Electronics, FCT-ZTF, UPV-EHU, 48080 Bilbao, Spain}

\author{E. E. Krasovskii}
\affiliation{Donostia International Physics Center (DIPC), Paseo Manuel de Lardizabal 4, 20018 Donostia/San Sebasti\'{a}n,  Basque Country, Spain}
\affiliation{Departamento de Pol\'{i}meros y Materiales Avanzados: F\'{i}sica, Qu\'{i}mica y Tecnolog\'{i}a, Universidad del Pa\'{i}s Vasco/Euskal Herriko Unibertsitatea, 20080 Donostia/San Sebasti\'{a}n, Basque Country, Spain}
\affiliation{IKERBASQUE, Basque Foundation for Science, 48013 Bilbao, Basque Country, Spain}

\date{\today}

\begin{abstract}
A theoretical study is presented of the effect of an in-plane magnetic exchange field on the band structure of  centrosymmetric films of noble metals and topological insulators. Based on an {\it ab initio} relativistic \kp\ theory, a minimal effective model is developed that describes two coupled copies of a Rashba or Dirac electronic system residing at the opposite surfaces of the film. The coupling leads to a structural gap at $\bar{\Gamma}$ and causes an exotic redistribution of the spin density in the film when the exchange field is introduced. We apply the model to a nineteen-layer Au(111) film and to a five-quintuple-layer Sb$_2$Te$_3$ film. We demonstrate that at each film surface the exchange field induces spectrum distortions similar to those known for Rashba or Dirac surface states with an important difference due to the coupling: At some energies, one branch of the state loses its counterpart with the oppositely directed group velocity. This suggests that a large-angle electron scattering between the film surfaces through the interior of the film is dominant or even the only possible for such energies. The spin-density redistribution accompanying the loss of the counterpart favors this scattering channel.
\end{abstract}

\maketitle

\section{Introduction}

Two-dimensional (2D) ferromagnets with strong spin-orbit interaction (SOI) possess a variety of spin magnetic properties relevant for spintronics, including anomalous Hall effect~\cite{Nagaosa10}, spin Hall effect in ferromagnets~\cite{Sinova15}, quantum anomalous Hall effect~\cite{Yu10, Wang15, Hou19}, anisotropic magnetoresistance, and planar Hall effects of various origins~\cite{Scharf16, Taskin17, Zheng20, Rao21}. The time-reversal symmetry breaking accompanied by the spin-momentum locking of the 2D states leads to the antisymmetric spin filtering~\cite{Streda03} and causes spin-orbit torques~\cite{Manchon19} and spin swapping~\cite{Saidaoui16}. Although these problems are, in principle, theoretically accessible with {\it ab initio} methods~\cite{Gradhand12, Lowitzer11, Freimuth14}, effective models~\cite{Streda03, Yu10, Wang15, Scharf16, Wang16, Hou19, Zheng20, Thalmeier20, Rao21} are indispensable, as they provide a greater freedom of modeling.

The majority of studies have addressed crystal surfaces of sufficiently thick films, in which the interaction between the two 2D systems on the opposite surfaces can be neglected. However, at metallic surfaces the spin-orbit-split surface states often energetically overlap with the bulk bands, which reduces the lifetime of the 2D carriers. This draws the attention to ultra-thin films, in which the interaction between the two surfaces cannot be neglected, and calls for the development of effective \kp\ models capable of describing such systems. So far in the \kp\ modeling of the Rashba-split or topological surface states the interaction between the states at the opposite surfaces has been described by a tunneling parameter having the same structure as the Zeeman interaction~\cite{Yu10, Wang15, Hou19, Thalmeier20, Rao21}, which is sufficient to mimic the structural gap and, consequently, a finite effective mass. However, in real materials the interplay between the structural splitting and spin polarization due to SOI may be rather complicated, with a nonuniform spin density distribution at each of the surfaces (see, e.g., the studies of the overlayers, Refs.~\cite{Shikin13} and \cite{Munoz_PRB_2014}), which is neglected in the simplified models. Here, we develop a relativistic effective \kp\ model that includes SOI, magnetic exchange interaction, and the spatial overlap between the 2D states---the precursors of the surface states---at the {\it ab initio} level. We apply the model to the study of the effect of the in-plane magnetization on the band structure of centrosymmetric films of noble metals and three-dimensional (3D) topological insulators.

Our proof-of-principle calculation shows that the structural gap presents new advantages for spin manipulation and scattering-channel engineering at the nanoscale.  We start with an {\it ab initio} relativistic \kp\ theory~\cite{Nechaev_PRBR_2016, Nechaev_PRB_2018, Nechaev_PRB_2019, Nechaev_PRB_2020} that generates effective Hamiltonians of a desired size and provides a reliable treatment of spin. This enables a predictive analysis of the effect of exchange magnetic interaction in accord with experimental observations~\cite{Susanne2019, Usachov_PRL_2020}. We will consider one representative of each class of materials: a nineteen-layer Au(111) film and a five-quintuple-layer Sb$_2$Te$_3$ film. A four-band Hamiltonian generated for these films is presented in a surface-resolved basis so that the resulting \kp\ model can be easily decomposed into two copies of a Rashba or Dirac electronic system and their interaction accurately described up to third order in $\mathbf{k}$. A distinctive and novel feature of the present {\it ab initio} \kp\ theory is that it makes no assumptions on the spatial structure of the wave functions, which allows us to consistently introduce the depth dimension into the spin-structure analysis.

We consider the simultaneous action of the SOI and an in-plane exchange field on the 2D states localized at the opposite surfaces and reveal that the interaction between the states plays a crucial role in the behavior of spin and restricts the scattering phase space for these states. In the absence of the exchange field, the SOI splits the dispersion of the surface-state precursor into two branches, with each branch giving rise to a closed constant energy contour at each of the surfaces. However, in the presence of the exchange field, for certain energies, one contour may be torn between the two surfaces, i.e., an open arc occurs at one surface, and its counterpart with the opposite group velocity is at the other one. The resulting very specific shape and spatial spin structure of the constant energy contours constrain the large-angle scattering so that it is necessarily accompanied by a jump to the opposite surface.

\section{Computational details}

The \textit{ab initio} band structure of the films is obtained in the repeated-slab model with the extended linear augmented plane waves method ~\cite{Krasovskii_PRB_1997} using the full potential scheme of Ref.~\cite{Krasovskii_PRB_1999} within the local density approximation (LDA). The spin-orbit interaction was treated by a second variation method~\cite{Koelling_1977}. The noble metal and topological insulator films are represented, respectively, by the bulk-truncated centrosymmetric nineteen-layer slab of Au(111) and five-quintuple-layer (QL) slab of Sb$_2$Te$_3$ (both films have space group $P\bar{3}m1$, no.~164). For Sb$_2$Te$_3$, the experimental crystal lattice parameters were taken from Ref.~\cite{Wyckoff_RWG} with the LDA relaxed atomic positions of Ref.~\cite{Nechaev_PRB_2015_SBTE}. The experimental lattice parameter of gold was taken from Ref.~\cite{Maeland_CJP_1964}. The films are thick enough to simulate the classical Rashba or Dirac surface state, but, at the same time, there is a tangible splitting of the surface state at $\bar{\Gamma}$: being an eigenfunction of a centrosymmetric slab Hamiltonian, the surface state is represented by two doubly degenerate slab levels $E_1$ and $E_2$ separated by a structural gap of a few meV, $\Delta=E_2-E_1$. This means that at $\bar{\Gamma}$ the Rashba or Dirac surface states form two Kramers-degenerate pairs with the spinor wave functions $|\Psi_{1\mu}\rangle$ and $|\Psi_{2\mu}\rangle$, Figs.~\ref{fig1}(a) and \ref{fig1}(b).

\section{Minimal effective model}

We start with a four-band \kp\ model, choosing the \textit{ab initio} spinors  $|\Psi_{1\mu}\rangle$ and $|\Psi_{2\mu}\rangle$ as the basis functions, where the subscript $\mu=\uparrow$ or $\downarrow$ indicates the sign of the {\it on-site} expectation value of the $z$ component $\widehat{J}_z$ of the total angular momentum $\widehat{\mathbf{J}} = \widehat{\mathbf{L}} + \widehat{\mathbf{S}}$~\cite{Nechaev_PRBR_2016, Nechaev_PRB_2020}. With this basis set, we derive a four-band \kp\ Hamiltonian $H_{\rtm{\mathbf{kp}}}$ from the \textit{ab initio} relativistic \kp\ perturbation expansion carried out around the $\Gamma$ point up to the third order in $\mathbf{k}$~\cite{Nechaev_PRB_2020}. Next, we transfer to the new basis $|\Phi^{\pm}_{\mu}\rangle =\frac{1}{\sqrt{2}} \left[|\Psi_{1\mu}\rangle \pm |\Psi_{2\mu}\rangle\right]$~\cite{Nechaev_PRB_2018}, in which the four-band Hamiltonian reads
\begin{equation}\label{HamFilm4x4}
H_{\rtm{\mathbf{kp}}}\longrightarrow H^{\rtm{Film}}_{\rtm{\mathbf{kp}}}=\left(
\begin{array}{cc}
H_{\rtm{Surf}}^{+} & H_{\rtm{int}}  \\
H^{\dag}_{\rtm{int}} & H_{\rtm{Surf}}^{-}
\end{array}
\right).
\end{equation}
Here, $H_{\rtm{Surf}}^{\pm}=[\epsilon+Mk^2]\rtm{\mathbb{I}}_{2\times2}\pm H_{\rtm{R}}$ and the interaction term $H_{\rtm{int}}=[\Delta\epsilon+\Delta Mk^2+i\Delta W(k_+^3+k_-^3)]\rtm{\mathbb{I}}_{2\times2}$ with $k=\sqrt{k_x^2+k_y^2}$, $k_{\pm}=k_x\pm ik_y$, and $\widehat{\mathbf{x}}$ being the direction $\bar{\Gamma}$-$\bar{M}$. The well-known $2\times2$ Rashba term
\begin{equation}\label{Ham_rash}
H_{\rtm{R}}=\left(
\begin{array}{cc}
iW (k_+^3-k_-^3)    & i\alpha k_- \\
-i\alpha k_+  & -iW(k_+^3-k_-^3)
\end{array}
\right),
\end{equation}
is responsible for the out-of-plane and in-plane spin structure typical of hexagonal surfaces, see, e.g, Ref.~\cite{Nechaev_PRB_2019} and references therein. In Eq.~(\ref{Ham_rash}), the second-order-corrected Rashba parameter is $\alpha=\alpha^{(1)}+\alpha^{(3)}k^2$.

In our theory, a reliable treatment of spin~\cite{Nechaev_PRB_2018, Susanne2019, Usachov_PRL_2020} is realized by means of the spin matrix
\begin{equation}\label{SpinFilm4x4}
\rtm{\mathbf{S}}_{\rtm{\mathbf{kp}}}\longrightarrow \rtm{\mathbf{S}}^{\rtm{Film}}_{\rtm{\mathbf{kp}}}=\left(
\begin{array}{cc}
\rtm{\mathbf{S}} & \widetilde{\rtm{\mathbf{S}}}  \\
\widetilde{\rtm{\mathbf{S}}} & \rtm{\mathbf{S}}
\end{array}
\right),
\end{equation}
where $\rtm{\mathbf{S}}=(s^{\spr}\bm{\sigma}_{\spr}, s^{z}\sigma_z)$ and $\widetilde{\rtm{\mathbf{S}}}=(\Delta s^{\spr}\bm{\sigma}_{\spr}, \Delta s^{z}\sigma_z)$. The elements of the spin matrix $[\rtm{\mathbf{S}}^{\rtm{Film}}_{\rtm{\mathbf{kp}}}]^{\mu\tau}_{\nu\chi} = \langle\Phi^{\tau}_{\mu}|\bm{\sigma}|\Phi^{\chi}_{\nu}\rangle$, where $\tau$ and $\chi$ are $+$ or $-$, enter the expression for the spin expectation value
\begin{equation}\label{modelRealSpin}
\langle \mathbf{S}_{\mathbf{k}\lambda}\rangle = \frac{1}{2} \langle \widetilde{\Phi}^{\lambda}_{\mathbf{k}}|\bm{\sigma}|\widetilde{\Phi}^{\lambda}_{\mathbf{k}}\rangle
= \frac{1}{2}\sum\limits_{\mu\tau \nu\chi} C_{{\mathbf{k}}\mu\tau}^{\lambda\ast}C_{{\mathbf{k}}\nu\chi}^{\lambda} \left[\rtm{\mathbf{S}}^{\rtm{Film}}_{\rtm{\mathbf{kp}}}\right]^{\mu\tau}_{\nu\chi}
\end{equation}
in the model state $|\widetilde{\Phi}^{\lambda}_{\mathbf{k}}\rangle = \sum\limits_{\mu\tau}C_{\mathbf{k}\mu\tau}^{\lambda} |\Phi^{\tau}_{\mu}\rangle$ of the reduced Hilbert space of the Hamiltonian~(\ref{HamFilm4x4}). The four-dimensional vectors $\mathbf{C}^{\lambda}_{\mathbf{k}}$ diagonalize this Hamiltonian  $H^{\rtm{Film}}_{\rtm{\mathbf{kp}}} \mathbf{C}^{\lambda}_{\mathbf{k}} = E^{\lambda}_{\mathbf{k}} \mathbf{C}^{\lambda}_{\mathbf{k}}$. Below, for simplicity, the in-plane component of the spin $\langle \mathbf{S}_{\mathbf{k}\lambda}\rangle$ will be referred to as $\mathbf{S}_{\spr}$.

\begin{figure}[tbp]
\centering
 \includegraphics[width=\columnwidth]{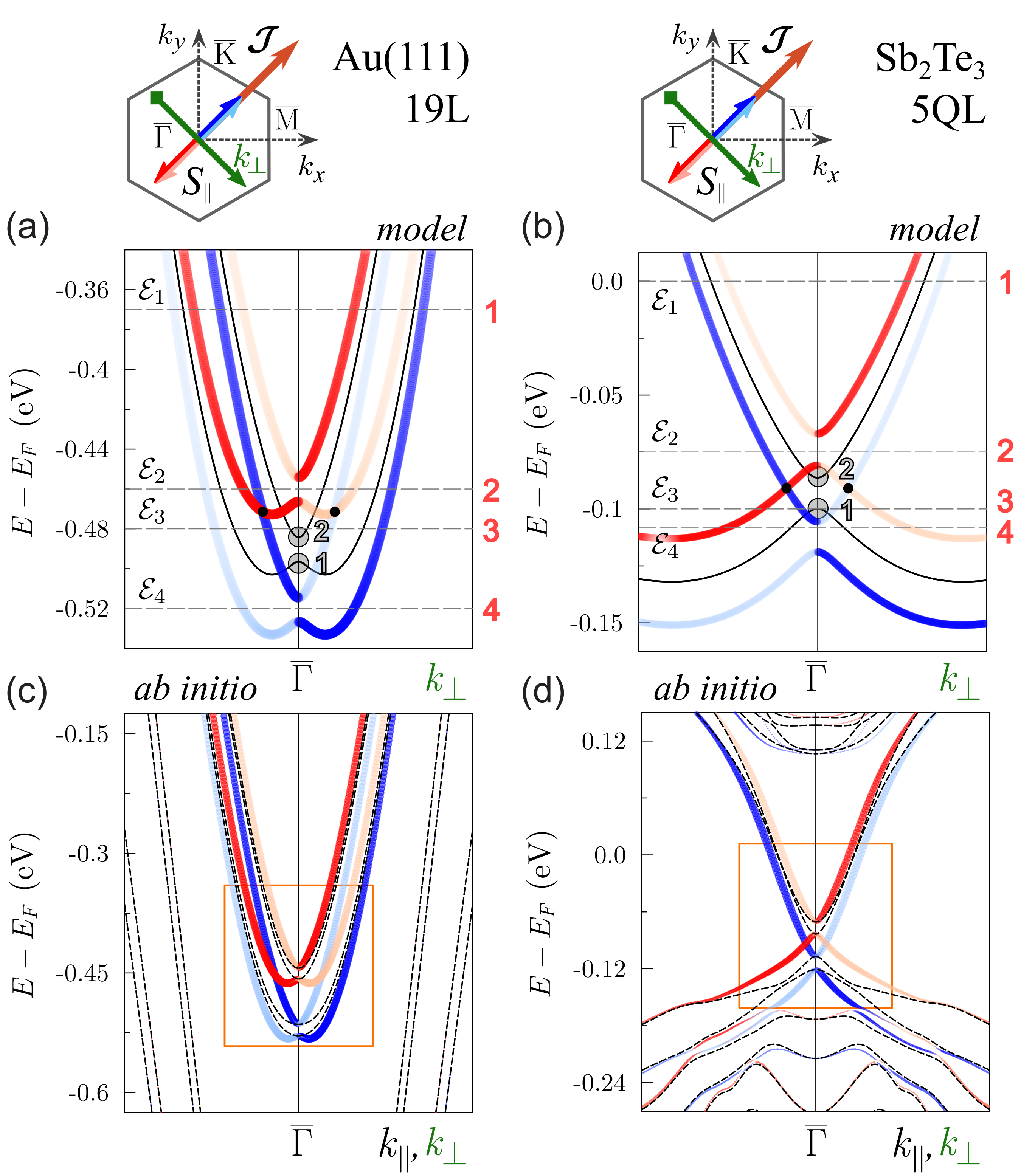}
\caption{Band structure of the nineteen-layer Au(111) film (a) and the five-QL Sb$_2$Te$_3$ film (b) by the four-band \kp\ Hamiltonian shown by black lines for the exchange parameter $\bm{\mathcal{J}} = 0$ and by fat bands for $\bm{\mathcal{J}} = \mathcal{J} (\widehat{\mathbf{x}} + \widehat{\mathbf{y}}) / \sqrt{2}$ with $\mathcal{J}=30$~meV for $\mathbf{k} = k_{\bot} (\widehat{\mathbf{x}} - \widehat{\mathbf{y}})/ \sqrt{2}$ perpendicular to the magnetization. The sign of the in-plane spin projection perpendicular to $\mathbf{k}$ is shown by color: Bright colors (red and blue) are used for the upper surface and pale colors for the lower surface. In graphs (a) and (b), the gray numbers mark the basis states $|\Psi_{1}\rangle$ and $|\Psi_{2}\rangle$, while the red ones label the energies $\mathcal{E}_1$, $\mathcal{E}_2$, $\mathcal{E}_3$, and $\mathcal{E}_4$ at which the constant energy contours shown in Figs.~\ref{fig2} and \ref{fig3} are calculated. Graphs (c) and (d) show, respectively, the \ai\ band structure of the Au(111) and Sb$_2$Te$_3$ film (see text). The bands for the in-plane exchange field are represented by fat bands for $\mathbf{k}\perp \bm{\mathcal{J}}$ ($k_{\bot}$-axis) and by dashed black lines for $\mathbf{k}\| \bm{\mathcal{J}}$ ($k_{\|}$-axis). Here, apart from the sign of the in-plane spin projection the fat bands reflect the localization of the states in three uppermost (bright colors) and lowermost (pale colors) atomic layers of the films. Orange rectangles outline the energy-momentum regions of graphs (a) and (b).}
\label{fig1}
\end{figure}

\begin{table}[b]
\caption{\label{tab:table1} Parameters of the Hamiltonian~(\ref{HamFilm4x4}) for Au(111) nineteen-layer slab with the lattice parameter $a=5.4495$~a.u. and Sb$_2$Te$_3$ five-QL slabs with $a=8.0312$~a.u. in Rydberg atomic units (except $\epsilon$ and $\Delta \epsilon$ given in eV).}
\begin{ruledtabular}
\begin{tabular}{ldd}
                                                      & \multicolumn{1}{c}{Au(111) 19L} & \multicolumn{1}{c}{Sb$_2$Te$_3$ 5QL} \\
\hline
$\epsilon$                                   &  -0.490               &    -0.093   \\
$\alpha^{(1)}$                           & -0.134             & -0.271   \\
$\alpha^{(3)}$                           & 10.14              & -33.33   \\
$W$                                              &  0.10              &  -52.71   \\
$M$                                 &   5.06              &   7.37   \\
$s^{\spr}$                                   &   0.98              &   0.62   \\
$ s^{z}$                                        &   0.96              &   0.25   \\
\hline
$\Delta \epsilon$                      &   -0.006                   &  -0.007      \\
$\Delta M$                                &  -0.07               &  -0.80   \\
$\Delta W$                                 &  -0.03              &  -1.71   \\
$\Delta s^{\spr}$                      &   0.00              &   0.00   \\
$\Delta s^{z}$                            &  0.00               &  -0.01
\end{tabular}
\end{ruledtabular}
\end{table}

The microscopically obtained parameters in Eqs.~(\ref{HamFilm4x4}) and (\ref{SpinFilm4x4}) are listed in Table~\ref{tab:table1}. The eigenvalues of the Hamiltonian~(\ref{HamFilm4x4}) obtained with these parameters is shown in Figs.~\ref{fig1}(a) and \ref{fig1}(b) by black solid lines. The spectra are represented by doubly degenerate bands with the characteristic Rashba- or Dirac-like behavior and exhibit the structural gap at $\bar{\Gamma}$ due to the coupling between two copies of the Rashba or Dirac electronic systems residing at the opposite surfaces of a film. Note that our \textit{ab initio} relativistic \kp\  theory provides an accurate description of this coupling by the term $H_{\rtm{int}}$ up to the third order in $\mathbf{k}$. For further purposes, we define the depth parameter for a given model state $|\widetilde{\Phi}^{\lambda}_{\mathbf{k}}\rangle$ as
\begin{equation}\label{Depth_film}
D^{\lambda}_{\mathbf{k}} = \sum\limits_{\mu}\left(|C_{\mathbf{k}\mu+}^{\lambda}|^2 - |C_{\mathbf{k}\mu-}^{\lambda}|^2\right).
\end{equation}
This parameter varies from $-1$ (the upper film surface) to $1$ (the lower film surface) and adds a new dimension to our analysis to characterize the spatial localization of the model states in the films.

Now, we include the exchange field into the magnetic Hamiltonian $H^{\rtm{M}}_{\rtm{\mathbf{kp}}} = H^{\rtm{Film}}_{\rtm{\mathbf{kp}}}+ H_{\rtm{EX}}$ as an exchange term $H_{\rtm{EX}} = -\bm{\mathcal{J}} \cdot$ $\rtm{\mathbf{S}}^{\rtm{Film}}_{\rtm{\mathbf{kp}}}$, where $\bm{\mathcal{J}}=J_{\rtm{ex}}\mathbf{M}$ is a tunable parameter allowing for the magnetic exchange interaction of strength $J_{\rtm{ex}}$ with a magnetization $\mathbf{M}$, when, e.g., the film is brought into contact with a magnetic material or contains ferromagnetically ordered magnetic (spin) moments of doped transition-metal atoms. In the present study, we consider an in-plane magnetization with $\bm{\mathcal{J}} = \mathcal{J} (\widehat{\mathbf{x}} + \widehat{\mathbf{y}}) / \sqrt{2}$ and $\mathcal{J}=30$~meV.

For the films we study, the in-plane components of the non-diagonal block $\widetilde{\rtm{\mathbf{S}}}$ of the spin matrix~(\ref{SpinFilm4x4}) are zero, and, therefore, the effect of the in-plane exchange field is described by the block diagonal matrix $H_{\rtm{EX}}$ whose elements are added to the elements of $H_{\rtm{Surf}}^{\pm}$. This implies that in both the non-magnetic and magnetic phase the interaction between the surfaces is exclusively due to the term $H_{\rtm{int}}$, whose parameters are rather small for the chosen thicknesses of the films, Table~\ref{tab:table1}. With increasing the film thickness, these parameters become negligible, and we arrive at two uncoupled surfaces. Each of the surfaces is described by the spin matrix $\rtm{\mathbf{S}}$ and the Hamiltonian $H_{\rtm{Surf}}^{+}$  or $H_{\rtm{Surf}}^{-}$, which is in accord with the form of the two-band Hamiltonian constructed in Ref.~\cite{Fu_PRL_2009} by considering the $C_{3v}$ crystal symmetry and time-reversal symmetry only. Because it comes directly from our fully {\it ab initio} \kp\ perturbation approach, this Hamiltonian is the same both for the Dirac surface state of topologically nontrivial insulators and for the so-called Rashba-split surface state of noble metals. This can be viewed as an \ai\ confirmation of the applicability of the two-band Hamiltonian of Ref.~\cite{Fu_PRL_2009} to topological surface states and the Rashba Hamiltonian of the semiconductor quantum well physics~\cite{Rashba_FTT_1959, Rashba_JETPL_1984} to the trivial surface states first suggested by LaShell \textit{et al.}~\cite{LaShell_PRL_1996}. Additionally, we note that for the Au(111) film the spin parameters $s^{\spr}$ and $s^{z}$ are almost unity, see Table~\ref{tab:table1}, and in this case one may associate the Pauli matrices generally used to represent the Hamiltonian with the observable spin.

Furthermore, Hamiltonian~(\ref{HamFilm4x4}) is an instructive \kp\ model for studying a phenomenon known as a hidden Rashba effect~\cite{Zhang_hidden_2014}, which here is associated with two inversion-symmetry related copies of the Rashba or Dirac system. One copy is described by $H_{\rtm{Surf}}^{+}$, and its inversion partner is represented by $H_{\rtm{Surf}}^{-}$. However, although the \kp\ model for the isolated copies is well known, their interaction has not been hitherto accurately treated within a relativistic \kp\ theory. Our \textit{ab initio} \kp\ perturbation expansion yields the interaction term $H_{\rtm{int}}$ up to third order in $\mathbf{k}$ in accord with the order of the spin-orbit term $H_{\rtm{R}}$ accounting for the splitting and spin structure of the Rashba or Dirac state. In this regard, it is also worth noting that besides a more accurate description of the dispersion of the surface-state precursors than in the classical (linear) Rashba and Dirac models, the third-order \kp\ model allows for a nonzero out-of-plane spin polarization $S_z$ permitted by symmetry, for the hexagonal warping of constant energy contours (CECs), and for the anisotropic \textbf{k}-dependent interaction between the states at the opposite film surfaces. As seen in Table~\ref{tab:table1}, the \ai\ values of the parameters $\alpha^{(3)}$, $W$, and $\Delta W$ for the Sb$_2$Te$_3$ film are substantially larger than those for the Au(111) film, and, therefore, the improvement over the classical models is more significant for the topological insulator film.

\section{In-plane exchange field effect}

\begin{figure*}[tbp]
\centering
 \includegraphics[width=\textwidth]{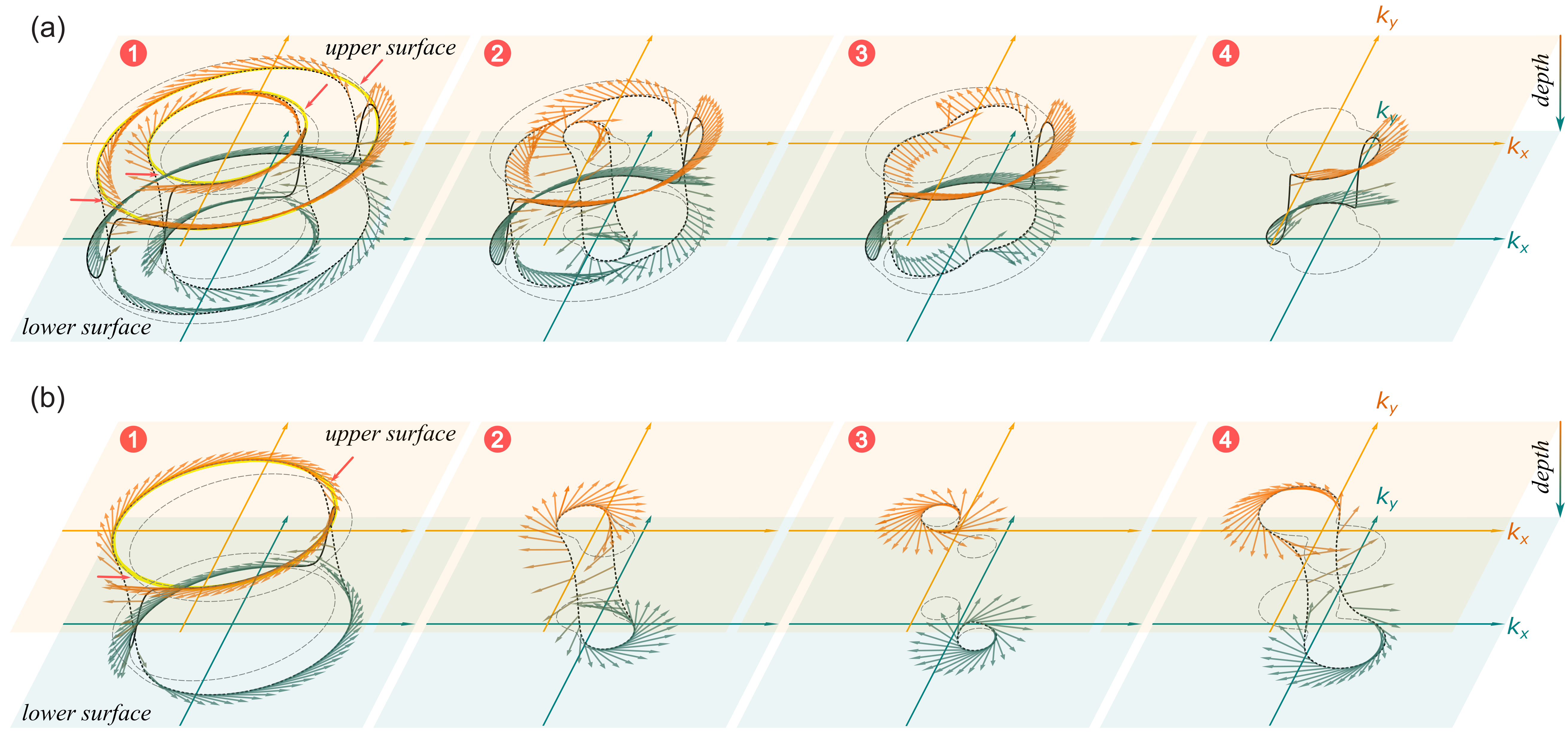}
\caption{Depth- and spin-resolved constant energy contours for the nineteen-layer Au(111) film (a) and the five-QL Sb$_2$Te$_3$ film (b) under the in-plane exchange field at the energies marked by red numbers in Figs.~\ref{fig1}(a) and \ref{fig1}(b), respectively. The in-plane spin is represented by arrows of the color changing from orange to teal according to the increase of the depth from the upper to the lower surface. The depth resolution manifests itself in the stretching of the contours (solid and dotted black lines) in the depth dimension determined by the contributions of the surface-related basis states $|\Phi^{\pm}\rangle$ to the model state of the four-band \kp\ Hamiltonian, Eq.~(\ref{Depth_film}). Gray dashed lines are the projection of the contours onto the surface planes. In graphs showing the contours at $\mathcal{E}=\mathcal{E}_1$, yellow circles are the contours for a semi-infinite film under the field.}
\label{fig2}
\end{figure*}

The band structures of the Au(111) and Sb$_2$Te$_3$ films in the in-plane exchange field by the magnetic \kp\ Hamiltonian $H^{\rtm{M}}_{\rtm{\mathbf{kp}}}$ with $\mathcal{J}=30$~meV are shown in Figs.~\ref{fig1}(a) and \ref{fig1}(b) by fat bands highlighting the sign of the in-plane spin projection onto the field direction for $\mathbf{k}\perp \mathbf{M}$. As seen in the figures, due to the exchange interaction the states with the in-plane spin $\mathbf{S}_{\spr}$ co-directional with the field (blue-shade bands) tend to decrease their energy, while the states with $\mathbf{S}_{\spr}$ opposite to the field (red-shade bands) acquire higher energy. The resulting red- and blue-shade bands resemble an ordinary Zeeman splitting of scalar-relativistic doubly degenerate bands, i.e., the zero-field band structures (black lines) shifted, respectively, up and down in energy.

To understand the effect of the SOI-induced spin structure, one should take into account the depth-localization of the states according to the $D^{\lambda}_{\mathbf{k}}$  parameter~(\ref{Depth_film}) given by the color shade. In this case, one can clearly distinguish two pairs of the branches localized at opposite surfaces---the bright red and blue branches of the upper surface and the pale ones of the lower surface. Each pair demonstrates the well-known modifications of the classical Rashba or Dirac states by an in-plane magnetic exchange field with the crossing points shifted away from $\bar{\Gamma}$. As seen in Figs.~\ref{fig1}(a) and \ref{fig1}(b), these crossing points (black points in the figures) are on opposite sides of $\bar{\Gamma}$ for the opposite surfaces due to the different sign of the SOI term $H_{\rtm{R}}$ in $H_{\rtm{Surf}}^{\pm}$. Note that the magnetic \kp\ calculations highly accurately reproduce the dispersion, spin polarization, and localization of the true bands near the $\bar{\Gamma}$ point, see Figs.~\ref{fig1}(c) and \ref{fig1}(d). These figures show the \ai\ band structures of the films with the in-plane magnetization realized by adding a uniform Zeeman field to the original LDA Hamiltonian as implemented in the FLEUR code~\cite{FLEUR}.

In the presence of the field, the structural gap due to the coupling does not disappear either in the model or in \ai\ spectra, and at  $\bar{\Gamma}$ it breaks each of the spin-split branches of the surface-state precursors, Fig.~\ref{fig1}. Owing to the gap at $\bar{\Gamma}$, for $\mathbf{k} \perp \mathbf{M}$ there is an energy interval where a branch of the split state at the upper or lower surface loses its counterpart with the opposite group velocity. In the Rashba system, this is the branch of the same color, implying the same in-plane spin projection, while in the Dirac system the lost counterpart is the branch of the other color, i.e., with the flipped $\mathbf{S}_{\spr}$.

A similar picture of the unpaired branches was observed in Ref.~\cite{Carbone_PRB_2016} for Ag$_2$Bi/Ag/Fe, but there it was caused by a spin-selective hybridization between the exchange-split quantum well states of a fifteen-layer Ag(111) film grown on the ferromagnetic substrate Fe(110) and the Rashba-split states of the surface alloy Ag$_2$Bi on the top of the silver film. In our case,  the exchange field does not act through an intermediary, but acts directly on the structurally split Rashba or Dirac states, Fig.~\ref{fig1}. A way to realize this in a real system is to sandwich the films between the magnetic layers of a van der Waals transition-metal or rare-earth trihalide (for example, for the topological insulator film it might be the ferromagnetic layer of GdI$_3$~\cite{You_PRB_2021}).

In order to examine the structural-gap effect over the whole $(k_x,k_y)$ plane, we employ our \kp\ theory to calculate spin-resolved CECs of the surface-state precursors around $\bar{\Gamma}$ at the energies indicated in Figs.~\ref{fig1}(a) and \ref{fig1}(b). The depth-resolved contours are presented in Fig.~\ref{fig2} as 3D curves, the vertical dimension being the depth defined by Eq.~(\ref{Depth_film}). As seen in the figure, these 3D CECs are strongly bent towards the vertical for $\mathbf{k}$ close to the field direction (the momentum polar angle $\varphi_{\mathbf{k}}\sim\pi/4$ and $\sim5\pi/4$), so one half of the contour lies on the upper and the other on the lower surface. Over the 3D CEC fragments that pass through the film the surface states are largely hybridized.

In the 2D projections of the 3D CECs onto the surface (dashed gray lines in Fig.~\ref{fig2}), the large hybridization manifests itself as the avoided crossing between the contours of the uncoupled surfaces, see red arrows for the case~1 in Figs.~\ref{fig2}(a) and \ref{fig2}(b). Indeed, a pair of the 2D projections (hereafter, 2D contours) can be easily recognized to be the exchange split contour, which is doubly degenerate in a nonmagnetic film. [For Au(111), there are two pairs: two inner and two outer 2D CECs.] In such a pair, one 3D CEC (solid black line) has two flat arcs that lie on the opposite surfaces and project onto one closed 2D CEC, and the arcs of the other one (dotted black line) project onto the other 2D CEC of the pair. In the plane, the arcs are disconnected around the avoided-crossing points, contrary to the case of the uncoupled surfaces, see yellow circles in Fig.~\ref{fig2}. (The avoided crossing comes from the structural splitting of the surface-state bands clearly seen for $\mathbf{k}\|\mathbf{M}$, see dashed black lines in Figs.~\ref{fig1}(c) and \ref{fig1}(d); for the uncoupled surfaces these bands are doubly degenerate along the $\mathbf{k}\|\mathbf{M}$.)

In some cases, in the flat arcs of the 3D CECs the familiar surface spin structure can be easily recognized (orange or teal arrows in Fig.~\ref{fig2}). For example, in the case 1 ($\mathcal{E}=\mathcal{E}_1$), each surface exhibits a pattern typical of a Rashba or Dirac system in an external in-plane exchange field: The exchange interaction shifts the contours perpendicular to the field and distorts the SOI-induced spin-momentum locking, causing anisotropy in the transport properties of the films. (Note a much greater impact of the exchange field on the in-plane spin structure of the Rashba than of the Dirac system, for which the linear Dirac model (not shown) gives an even smaller impact leaving the $\mathbf{S}_{\spr}$  pattern practically unaltered except for the vicinity of the avoided-crossing points.) Concerning the spin behavior along the whole 3D CECs, note that each pair of the 2D contours contains a dotted-line 3D CEC with the spin $\mathbf{S}_{\spr}$ rotating twice by $2\pi$ along the contour in the $(k_x,k_y)$ plane (double spin winding) and a solid-line 3D CEC along which $\mathbf{S}_{\spr}$ merely deviates from the field direction without a complete $2\pi$ rotation (zero spin winding). Further, we will refer to these 3D CECs as a non-trivial and trivial CEC, respectively.

We focus now on the CECs at lower energies (numbered by 2, 3, and 4) and start with the Au(111) film, Fig.~\ref{fig2}(a). As seen in Fig.~\ref{fig1}(a), for $\mathcal{E}=\mathcal{E}_2$ the red branch loses its counterpart with the opposite group velocity, so the trivial (solid line) 3D contour of the inner pair disappears. As a consequence, at each surface a certain $\mathbf{k}$-sector becomes unavailable for the elastic scattering. For $\mathcal{E}=\mathcal{E}_3$, only the blue branches are left, Fig.~\ref{fig1}(a), with one pair of the 3D CECs. Note that the dotted-line contour becomes trivial, with $\mathbf{S}_{\spr}$ mostly perpendicular to the field direction. At the same time, the trivial solid-line CEC is characterized by $\mathbf{S}_{\spr}$ gravitated towards $\bm{\mathcal{J}}$, Fig.~\ref{fig2}(a). This substantially reduces the phase space for scattering transitions that may occur at one film surface. Finally, for $\mathcal{E}=\mathcal{E}_4$ there remains only one trivial 3D CEC with a sizable spin projection onto the field direction, which practically forbids any large-angle scattering at a given film surface, thereby causing the electrons to ``leak'' through the interior of the film to the opposite surface.

In the five-QL Sb$_2$Te$_3$ film, at $\mathcal{E}=\mathcal{E}_2$ there is only one CEC, which is characterized by the double winding of the in-plane vector $\mathbf{S}_{\spr}$. For the flat CEC arcs, the spin structure is very similar to a Dirac surface state with the spin-momentum locking only slightly affected by the field. The influence of the field increases in the interior of the film, where $\mathbf{S}_{\spr}$ is mainly opposite to the magnetization. The presence of only one 3D CEC reduces by half the phase space at each surface, thereby making a large-angle scattering only possible by means of a ``leakage'' to the opposite surface. For $\mathcal{E}=\mathcal{E}_3$, we have two 3D CECs, but now they lie almost entirely on the opposite surfaces of the film as if the surfaces were uncoupled. Moreover, both CECs have the helical in-plane spin structure typical of the Dirac surface state with the single winding of the in-plane spin. This means that around the energy of the Dirac points located away from $\bar{\Gamma}$, see black points in Fig.~\ref{fig1}(b), there is an energy interval for which the band structure of the thin film in the in-plane exchange field is independent of whether or not the structural gap exists. In our case, this interval covers the gap of the zero-field spectrum of the film, thereby causing a drastic change of the film properties by turning on the field. Finally, for $\mathcal{E}=\mathcal{E}_4$, again, there is only one CEC with the spin structure of inverted helicity with respect to that of the case~2 with $\mathbf{S}_{\spr}$ mostly directed along the field in the interior of the film. Note also a more pronounced impact of the field on the in-plane spin structure here.

\begin{figure}[tbp]
\centering
 \includegraphics[width=\columnwidth]{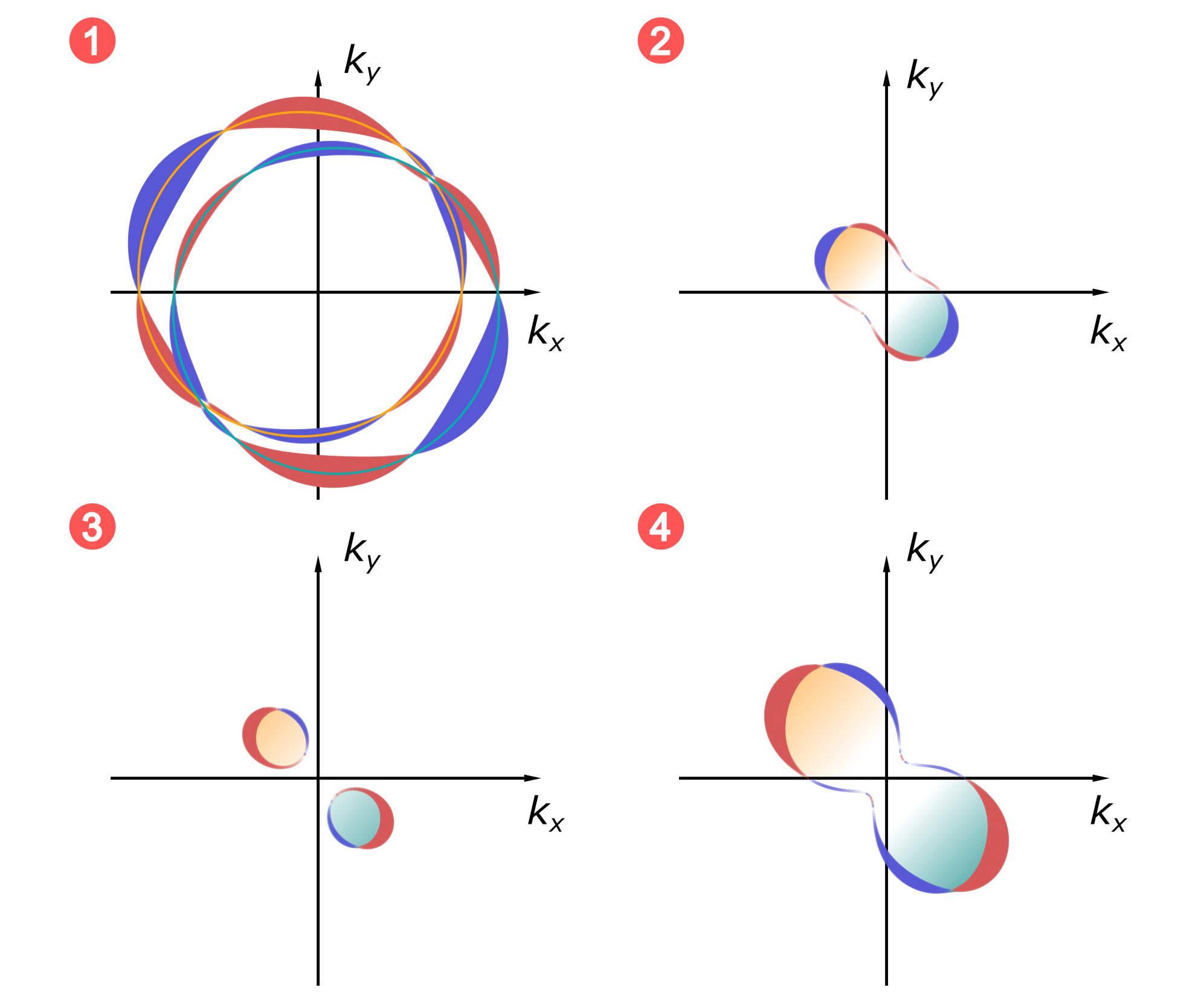}
\caption{Spin-resolved depth-integrated (``top view'') contours of the five-QL Sb$_2$Te$_3$ film at the energies marked in Fig.~\ref{fig1}(b). The expectation value of $S_z$ is shown by red ($S_z>0$) and blue ($S_z<0$) fat segments. Orange and teal circles in graph~1 and shadings in graphs~2--4 highlight the CECs segments related to the upper and lower surface, respectively.}
\label{fig3}
\end{figure}

We now turn to the spin $z$ polarization of the CECs of the five-QL Sb$_2$Te$_3$ film under the in-plane external exchange field. [In the Au(111) film the $S_z$ expectation value along the CECs is negligibly small at the energies we consider.] Figure~\ref{fig3} shows a ``top view'' of the spin-$z$ resolved CECs; the largest spin-$z$ projection along the CECs ($|S_z|\sim0.02$) is seen to occur at $\mathcal{E}=\mathcal{E}_1$.  In this case, the $S_z$ pattern with the specific motif inherent in the hexagonal structures resembles the pattern of an ordinary Dirac surface state. In Fig.~\ref{fig3}, the orange and teal circles are a guide for the eye to highlight the CECs segments related to the upper and lower surface, respectively. These circles are the CECs of a film with the uncoupled surfaces, and, therefore, they differ from the actual 2D projections of the 3D CECs in that they cross along the field direction. For lower energies ($\mathcal{E}=\mathcal{E}_2$, $\mathcal{E}_3$, and $\mathcal{E}_4$), this motif completely disappears in the $S_z$ pattern. Instead, there appear two rather large blue and red parts with a sizable $S_z$ on each surface-related sector of the 3D CECs (the pale orange and pale teal shading in Fig.~\ref{fig3}), while in the interior of the film (between the shaded areas) $S_z$ is negligible. For these energies, the CECs are characterized by a substantially smaller $S_z$ than in the $\mathcal{E}=\mathcal{E}_1$ case and by a slight field-induced imbalance between spin-up and spin-down (especially for $\mathcal{E}=\mathcal{E}_3$ and $\mathcal{E}_4$). The $\mathcal{E}=\mathcal{E}_3$ case demonstrates that, although the $\mathbf{S}_{\spr}$ pattern is similar to that of a Dirac surface state, the $S_z$ distribution is quite different. In all the cases, the $S_z$ distribution along the CECs gives rise to scattering transition constraints additional to those imposed by the in-plane spin.

\section{Conclusions}

To summarize, we have developed an effective \kp\ model in the Hilbert space of four basis states comprising both interacting surfaces of a thin film with strong spin-orbit coupling. We have applied the model to the description of the influence of the in-plane exchange field on the energy-momentum dispersion and spin-momentum locking of the surface-state precursors in a nineteen-layer Au(111) film and in a five-quintuple-layer Sb$_2$Te$_3$ film. The interaction between the states at the opposite surfaces makes the exchange-induced modifications strongly energy dependent: In the simplest case of energies well above the Rashba or Dirac point, a tangible difference from non-interacting surfaces is found only for \textbf{k} parallel to the field, where an avoided crossing appears between the constant energy contours of the surface states localized at the opposite surfaces. In the surface-resolved representation, this manifests itself in the partition of a contour at a given surface into two disconnected flat arcs. At lower energies, where the interaction between the surfaces leads to the structural gaps in the surface-state dispersion, one of the arcs disappears, so the large-angle scattering needs to be accompanied by a jump to the opposite surface across the interior of the film. Furthermore, we have located the energy intervals (one in the gold film and two in the topological-insulator film) where each surface hosts one arc only, so the large-angle electron scattering between the film surfaces is the only possible channel. This suggests a way to control the spin-selective transport properties of the films by manipulating the in-plane exchange field.

\begin{acknowledgments}
We acknowledge funding from the Department of Education of the Basque Government (Grant No.~IT1164-19) and the Spanish Ministry of Science, Innovation, and Universities (Project No.~PID2019-105488GB-I00).
\end{acknowledgments}

\end{document}